\documentclass[a4paper]{revtex4}
\usepackage{graphicx}
\usepackage{fancyhdr}
\usepackage{amsmath}
\usepackage{color}

\begin{document}

\title{NMSSM in TeV-scale mirage mediation}

%

\author{Tatsuo Kobayashi}
\affiliation{Department of Physics, Hokkaido University, Sapporo 060-0810, JAPAN}

\begin{abstract}
We study the next-to-minimal supersymmetric standard model 
with the TeV scale mirage mediation.
The 125 GeV Higgs boson mass is realized with ${\cal O}(10)\%$ tuning for 1.5 TeV gluino and 
1TeV stop masses.
This talk is based on Refs.\cite{Kobayashi:2012ee,ours}.
\end{abstract}

\maketitle

\thispagestyle{fancy}


\section{Introduction}
Supersymmetric extension of the standard model is one of interesting candidates 
for the physics beyond the standard model.
The LHC Run I did not find any superpartners, but put lower bounds 
for superpartner masses, e.g. about 1.5 TeV for the gluino mass and 700 GeV for the stop mass.

Within the framework of the 
minimal supersymmetric standard model (MSSM), 
the $Z$-boson mass, $m_Z$, is obtained as 
\begin{equation}
m^2_Z  \simeq - 2 m_{H_u}^2 + \frac{2}{\tan^2 \beta} m_{H_d}^2
  - 2\mu^2 \   .
\label{eq:mZ}
\end{equation}
Here, $m^2_{H_u}$ and $m^2_{H_d}$  are the soft supersymmetry (SUSY) breaking 
scalar mass squared of 
the up-sector and down-sector Higgs fields.
On the other hand, 
$\mu$ is the supersymmetric Higgs and higgsino mass and 
the corresponding superpotential term is written by
\begin{equation}
W_{\rm MSSM-Higgs} = \mu H_u H_d,
\end{equation}
where $H_u$ and $H_d$ are up-sector and down-sector Higgs superfields.
The gluino mass $M_3$ is dominant in radiative corrections on $m^2_{H_u}$ 
for many models, e.g. the constrained MSSM, and then 
we obtain $m^2_{H_u} \sim - M_3^2$.
Thus, if the gluino mass as well as the stop mass is of ${\cal O}(1)$TeV, 
we need fine-tuning among  $m^2_{H_u}$, $m^2_{H_d}$ and $\mu^2$ to 
derive the correct value of $m_Z$.
Indeed, the stop mass is required to be of ${\cal O}(1)$TeV or more 
in order to realize the 125 GeV Higgs mass.
Furthermore, in the MSSM, $\mu$ is the supersymmetric mass and 
there is no reason why $\mu$ is of the same order as soft SUSY breaking masses.
That is the so-called $\mu$-problem \cite{Kim:1983dt}.

The $\mu$-problem can be solved by extending the MSSM to the 
next-to-minimal supersymmetric standard model (NMSSM), where 
we add the singlet chiral multiplet $S$ \cite{Fayet:1974pd} 
(see for review e.g. \cite{Ellwanger:2009dp}). 
Then we can write the following superpotential terms including $S$,
\begin{equation}
W_{\rm NMSSM-Higgs} =  \lambda S H_u H_d + \frac{\kappa}{3}S^3.
\label{eq:W-NMSSM}
\end{equation}
We forbid the above $\mu$-term as well as the 
supersymmetric mass term of $S$ by assuming the $Z_3$ symmetry.
Thus, there is no supersymmetric mass terms.
By analyzing the scalar potential, we can determine the vacuum expectation value (VEV)
of $S$, which depends only on soft SUSY breaking parameters.
Then, we can obtain the effective $\mu$-term, $\mu = \lambda <S>$, 
which is of the same order as  other soft SUSY breaking parameters.
Even in the NMSSM, we face the fine-tuning problem when $m^2_{H_u} \sim - M_3^2$ 
and the gluino mass as well as the stop mass is of ${\cal O}(1)$TeV.

The concrete behavior of radiative corrections on $m^2_{H_u}$ as well as other soft masses 
depends on explicit spectrum of superpartners, that is, the mediation mechanism 
of SUSY breaking.
The mirage mediation is one of interesting mediation mechanisms \cite{Choi:2004sx,Choi:2005uz,Endo:2005uy}, and it is 
a mixture between anomaly mediation and modulus mediation.
In the mirage mediation, the radiative corrections and anomaly mediation cancel each other 
at a certain energy scale, where the SUSY spectrum appears as the 
 pure modulus mediation.
Such an energy scale is called the mirage scale.
The TeV scale mirage mediation sets this energy scale around TeV scale.
Then, $m^2_{H_u}$ has no large radiative corrections due to $M_3$.
Indeed, it was pointed out that the TeV scale mirage mediation can ameliorate 
the fine tuning problem in the MSSM \cite{Choi:2005hd,Kitano:2005wc,Choi:2006xb}.
(Non-universal gaugino masses with a certain ratio may be useful to 
ameliorate fine-tuning \cite{Abe:2007kf}.)

In this talk, we apply the TeV scale mirage mediation to the NMSSM 
in order to improve the fine-tuning problem with deriving the 125 GeV Higgs mass.

\section{TeV scale mirage mediation}

Here we give a brief review on the mirage  mediation \cite{Choi:2004sx}.
The mirage mediation is the mixture of the modulus mediation 
and the anomaly mediation with a certain ratio, 
which would be determined by the modulus stabilization mechanism 
and SUSY breaking mechanism.
In the mirage mediation, the gaugino masses are written by 
\begin{eqnarray}\label{eq:gaugino-mass}
M_a(M_{GUT}) = M_0 +\frac{m_{3/2}}{8 \pi^2}b_a g_a^2,
\end{eqnarray}
where $g_a$ and $b_a$ are the gauge couplings 
and their $\beta$ function coefficients, and $m_{3/2}$ denotes 
the gravitino mass.
We assume that the initial conditions of our 
SUSY breaking parameters are input at the GUT scale, 
$M_{GUT} =2\times 10^{16}$ GeV.
The first term, $M_0$, in the right hand side denotes 
the gaugino mass due to the pure modulus mediation, 
while the second term corresponds to the anomaly mediation 
contribution.
In addition, we can write the soft scalar masses $m_i$ of 
matter fields $\phi^i$  and 
the so-called $A$-terms of $\phi^i \phi^j \phi^k$ corresponding to the
Yukawa couplings $y_{ijk}$ as 
\begin{eqnarray}\label{eq:A-m}
A_{ijk}(M_{GUT}) &=& a_{ijk}M_0 - (\gamma_i + \gamma_j + \gamma_k)\frac{m_{3/2}}{8\pi^2}, \nonumber \\
m_i^2(M_{GUT}) &=& c_i M_0^2 - \dot{\gamma_i}(\frac{m_{3/2}}{8\pi^2})^2
				- \frac{m_{3/2}}{8\pi^2} M_0 \theta_i,
\end{eqnarray}
where
\begin{eqnarray}
\gamma_i = 2\sum_a g_a^2 C_2^a(\phi^i) - \frac{1}{2} \sum_{jk} |y_{ijk}|^2, \quad
\theta_i = 4\sum_a g_a^2 C_2^a(\phi^i) - \sum_{jk} a_{ijk} |y_{ijk}|^2, \quad 
\dot{\gamma_i} = 8\pi^2 \frac{d\gamma_i}{d \ln \mu_R}.
\end{eqnarray}
Here, $\gamma_i$ denotes the anomalous dimensions of $\phi^i$ and 
$C_2^a(\phi^i)$ denotes the quadratic Casimir 
corresponding to the representation of the matter field $\phi^i$.
In the right hand side, $a_{ijk}M_0$  and $c_i M_0^2$ 
denote the A-term and soft scalar mass squared due to 
the pure modulus mediation.
These coefficients, $a_{ijk}$ and $c_i$, are determined by 
modulus-dependence of the K\"ahler metric of $\phi^i$, 
 $\phi^j$ and $\phi^k$ as well as Yukawa couplings.
Indeed, by using the tree-level K\"ahler metric, 
the coefficient $c_i$ is explicitly calculated as a fractional number 
such as $0,1,1/2,1/3$.
We would have ${\cal O}(1/8\pi^2)$ of corrections on 
$c_i$ due to the one-loop corrections on the K\"ahler metric.
Such a correction would be important when $c_i =0 $, 
but  that is model-dependent.
Here, we consider the case with 
\begin{equation}
a_{ijk} = c_i + c_j +c_k.
\end{equation}

It is convenient to use the following parameter \cite{Choi:2005uz},
\begin{equation}
\alpha \equiv \frac{m_{3/2}}{M_0 \ln(M_{pl}/m_{3/2})},
\end{equation}
to represent the ratio of the anomaly mediation 
to the modulus mediation. Here $M_{pl}$ is the reduced Planck scale.

The mirage mediation has a very important energy scale, that is, 
the mirage scale defined by, 
\begin{equation}
M_{\rm mir} = \frac{M_{GUT}}{(M_{pl}/m_{3/2})^{\alpha/2}} .
\end{equation}
The above spectrum of the gaugino masses at $M_{GUT}$ leads 
to  \cite{Choi:2005uz},
\begin{equation}\label{eq:gaugino-mir}
M_a(M_{\rm mir}) = M_0,
\end{equation}
at the mirage scale.
That is, the anomaly mediation contributions 
and the radiative corrections cancel each other, 
and the pure modulus mediation appears at the mirage scale.
Furthermore, the $A$-terms and the scalar mass squared
also satisfy 
\begin{equation}\label{eq:A-m-mir}
A_{ijk}(M_{\rm mir}) = (c_i+c_j+c_k)M_0, \qquad m^2_i(M_{\rm mir}) =c_iM_0^2,
\end{equation}
if the corresponding Yukawa couplings are small enough or 
if the following conditions are satisfied, 
\begin{equation}\label{eq:mir-condition}
a_{ijk}=c_i+c_j+c_k=1,
\end{equation}
for non-vanishing Yukawa couplings, $y_{ijk}$ \cite{Choi:2005uz}.

When $\alpha =2$, the mirage scale $M_{\rm mir}$ is 
around $1$ TeV.
Then, the above spectrum (\ref{eq:gaugino-mir}) and 
(\ref{eq:A-m-mir}) is obtained around the TeV scale.
That is the TeV scale mirage mediation scenario.
In particular, there would appear a large gap 
between $M_0$ and the scalar mass $m_i$ with $c_i \approx 0$.
We will apply the TeV scale mirage scenario to the NMSSM in the next section.

\section{NMSSM in TeV scale mirage mediation}

\subsection{NMSSM}

Here, we briefly review on the NMSSM, in particular its Higgs sector.
We extend the MSSM by adding a singlet chiral multiplet $S$ 
and imposing a $Z_3$ symmetry.
Then, the superpotential of the Higgs sector is written as Eq.~(\ref{eq:W-NMSSM}).
Here and hereafter, for $S$, $H_u$ and $H_d$ we use the convention that 
the superfield and its lowest component are 
denoted by the same letter.

The following soft SUSY breaking terms in the Higgs sector are 
induced, 
\begin{equation}
V_{\rm soft}= m^2_{H_u}|H_u|^2 + m^2_{H_d}|H_d|^2 
+m^2_S |S|^2-\lambda A_\lambda SH_uH_d + \frac{\kappa}{3}A_\kappa S^3+ h.c.
\end{equation}
Then, the scalar potential of the neutral Higgs fields is 
given as 
\begin{eqnarray}
V &=& \lambda^2|S|^2(|H^0_d|^2+|H^0_u|^2)+|\kappa S^2- \lambda
H^0_dH^0_u|^2 +V_D \nonumber \\
& & +m^2_{H_u}|H_u|^2 + m^2_{H_d}|H_d|^2 
+m^2_S |S|^2-\lambda A_\lambda SH_uH_d + \frac{\kappa}{3}A_\kappa S^3+ h.c.,
\end{eqnarray}
with
\begin{equation}
V_D = \frac18 (g^2_1 + g^2_2)(|H^0_d|^2 - |H^0_u|^2)^2,
\end{equation}
where $g_1$ and $g_2$ denote the gauge couplings of U(1)$_{\rm Y}$  
and SU(2).
Similarly, there appear the soft SUSY breaking terms 
including squarks and sleptons as well as gaugino masses.
These are the same as those in the MSSM.

The minimum of the Higgs potential is obtained by analyzing 
the stationary conditions of the Higgs potential,
\begin{equation}
\frac{\partial V}{\partial H^0_d} =\frac{\partial V}{\partial H^0_u} = \frac{\partial V}{\partial S} =0.
\label{eq:1}
\end{equation}
Here, we denote VEVs as 
\begin{equation}
v^2 = \langle |H^0_d|^2 \rangle + \langle |H^0_u|^2 \rangle, \qquad \tan \beta = \frac{\langle H^0_u \rangle }{\langle H^0_d \rangle}, 
\qquad s = \langle S \rangle.
\end{equation}
Using the above stationary conditions, we obtain the $Z$ boson mass $m_Z^2=\frac12 g^2v^2$ as
\begin{equation}
m_Z^2 = \frac{1 - \cos 2\beta}{\cos 2\beta} m_{H_u}^2 - \frac{1 + \cos 2\beta}{\cos 2\beta} m_{H_d}^2
  - 2\mu^2, 
\end{equation}
where $\mu = \lambda s$.
For $\tan\beta \gg 1$, this equation becomes Eq.~(\ref{eq:mZ}).
That is, 
this relation is the same as the one in the MSSM.
Thus, the natural values of $|m_{H_u}|$ and $|\mu|$ would be 
of ${\cal O}(100)$ GeV.
Furthermore, the natural value of $|m_{H_d}|/\tan \beta$ would be 
of  ${\cal O}(100)$ GeV or smaller.
Alternatively, $|\mu|$ and $|m_{H_d}|/\tan \beta$ could be larger 
than ${\cal O}(100)$ GeV when 
$\mu^2$ and $m_{H_d}^2/\tan^2 \beta$ are canceled each other 
in the above relation at a certain level.
Even in such a case, $|m_{H_u}|$ would be naturally of ${\cal O}(100)$ GeV.
On the other hand, other sfermion masses as well as gaugino 
masses must be heavy as the recent LHC results suggested.
To realize such a spectrum, we apply the 
TeV scale mirage mediation in the next section, 
where we take $c_{H_u}=0$ to realize a suppressed value of 
$|m_{H_u}|$ compared with $M_0$.

\subsection{TeV scale mirage mediation}

Here, we apply the TeV scalar mirage mediation scenario 
to the NMSSM and 
study its phenomenological aspects.
Soft SUSY breaking terms are obtained 
through the generic formulas (\ref{eq:gaugino-mass}) 
and (\ref{eq:A-m}) with taking $\alpha =2$.
We concentrate on the Higgs sector as well as gauginos and stops.

A concrete model in the mirage mediation is fixed by choosing $c_i$.
We consider the following values of $c_i$ \cite{Kobayashi:2012ee},
\begin{equation}\label{eq:ci}
c_{H_d}=1, \qquad  c_{H_u}=0, \qquad c_S = 0, \qquad c_{t_L}=c_{t_R}=\frac12, 
\end{equation}
up to one-loop corrections 
for $H_d$, $H_u$, $S$, and left and right-handed (s)top fields, respectively.
Then, the soft parameters due to only modulus mediation 
contribution are given by 
\begin{eqnarray}
& & (A_t)_{\rm modulus}=(A_\lambda )_{\rm modulus}=M_0, \qquad (A_\kappa )_{\rm modulus} =0, 
 (m^2_{H_d})_{\rm modulus}=M^2_0, \nonumber \\ 
& & (m^2_{\tilde t_L})_{\rm modulus} =
(m^2_{\tilde t_R})_{\rm modulus} =\frac12 M^2_0,  \qquad 
  (m^2_{H_u})_{\rm modulus }=(m^2_S)_{\rm modulus}=0,   
\end{eqnarray}
up to one-loop corrections.
The above assignment of $c_i$ (\ref{eq:ci})
satisfies the condition, (\ref{eq:mir-condition}) for 
the top Yukawa coupling and the coupling $\lambda$, but 
not for the coupling $\kappa$.
However, we do not consider a large value of $\kappa$ 
to avoid the blow-up of $\kappa$ and $\lambda$.
Thus, we obtain the following values,
\begin{eqnarray}\label{eq:soft-at-mirage-1}
&  A_t \approx A_\lambda \approx M_0, \qquad 
 m^2_{H_d} \approx M^2_0, \qquad m^2_{\tilde t_L} \approx m^2_{\tilde t_R} \approx \frac12 M^2_0, 
\end{eqnarray}
up to ${\cal O}(\kappa^2/8\pi^2)$ at the TeV scale.
Similarly, at the TeV scale we can obtain 
\begin{equation}\label{eq:soft-at-mirage-2}
m_{H_u}^2 \approx 0, \qquad 
m_{S}^2 \approx 0, \qquad
|A_\kappa|^2 \approx 0, 
\end{equation}
up to ${\cal {O}}(M_0/8\pi^2)$.
That is, values of $|A_\kappa|^2$, $m_{H_u}^2$ and 
$m_{S}^2$  are suppressed compared with 
$M_0^2$, and their explicit values depend on the one-loop corrections 
on the K\"ahler metric.
Thus, we use $A_\kappa$ 
as a free parameter, which must be small 
compared with $M_0$.
In addition, we determine the values of 
$m_{H_u}^2$, $m_S^2$ and $\mu~(=\lambda s)$ at the electroweak scale 
from the stationary conditions,  
(\ref{eq:1}), where we use the experimental value 
$m_Z = \frac{1}{\sqrt{2}} g v = 91.19$ GeV and $\tan \beta$ as a free parameter.

Through the above procedure, the parameters, $m^2_{H_u}$, $m^2_S$ and $\mu$, 
at the electroweak scale are expressed by $\tan\beta$, $m_{H_d}^2$, $A_\lambda$.
%
For $\tan\beta \gg {\rm max}( 1, \kappa/\lambda)$, these parameters are approximated as,
\begin{subequations}
\begin{align}
& \mu = \lambda \langle S \rangle \sim   \frac{m^2_{H_d}}{A_\lambda \tan \beta},  \label{eq:app-1}\\  
& m^2_S \sim -2 \left(\frac{\kappa}{\lambda}\right)^2\left(\frac{m_{H_d}^2}{A_\lambda\tan\beta}\right)^2-\left(\frac{\kappa}{\lambda}\right)A_\kappa \left(\frac{m_{H_d}^2}{A_\lambda\tan\beta}\right)+2\frac{\lambda^2}{g^2}\frac{A_\lambda^2}{m_{H_d}^2} m_Z^2, \label{eq:app-2}\\
& m_{H_u}^2 \sim  \frac{m^2_{H_d}}{\tan^2 \beta} -
\frac{m^4_{H_d}}{A^2_\lambda \tan^2 \beta}- \frac{m^2_Z}{2} .
\label{eq:app-3}
\end{align}
\label{eq:app}
\end{subequations}
See for more precise results Refs.\cite{Kobayashi:2012ee,ours}. 
When $\tan \beta ={\cal O}(10)$, 
the values of $\mu$, $|m_{H_u}|$ and $|m_S|$ are smaller than $M_0$ 
by the factor $\tan \beta$ because $m_{H_d} \simeq A_\lambda \simeq M_0$.
Thus, the values of $\mu$ and $|m_{H_u}|$ could be of ${\cal O}(100)$ GeV 
while the other masses of the superpartners are of ${\cal O}(M_0)={\cal O}(1)$ TeV.
Then, the fine-tuning problem can be ameliorated.
Furthermore, one can see that the first and the second terms in the last equation  
cancel each other for our choice of $c_i$. 
The next leading contributions are of ${\cal O}(m^2_{H_d}/\tan^4\beta)$ or
${\cal O}(m^2_{H_d}\mu/\tan^2\beta A_\lambda)$. 
Thus, $m^2_Z$ is almost determined by $m^2_{H_u}$ alone and
insensitive to the value of $\mu$. 
This means that actually $\tan\beta \approx 3$ is enough to obtain
the fine-tuning of $|\partial \ln m_Z^2/\partial \ln m_{H_u}^2|^{-1}=m_Z^2/2m_{H_u}^2={\cal O}(100)$\% for $M_0\approx 1$ TeV.
In this case, $\mu$ can be as heavy as ${\cal O}(400)$ GeV without introducing further fine-tuning.

\subsection{Higgs sector and fine-tuning}

\begin{figure}[h]
\begin{center}
\begin{tabular}{l @{\hspace{10mm}} r}
\includegraphics[height=45mm]{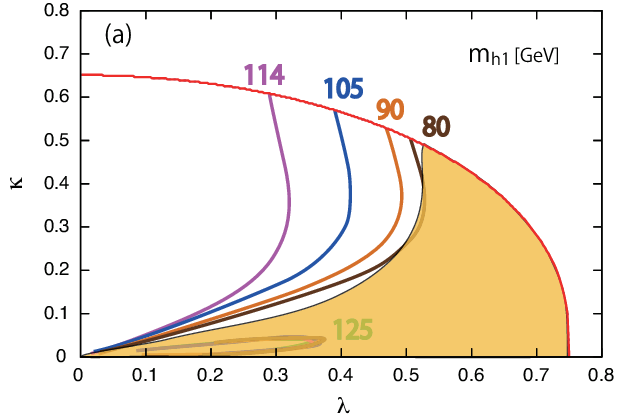} &
\includegraphics[height=45mm]{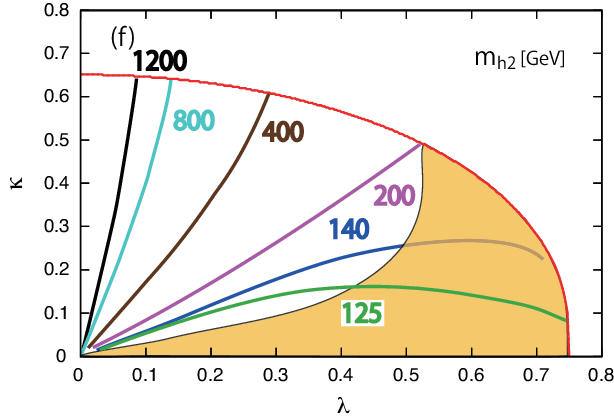} \\
\end{tabular}
\end{center}
\caption{The CP-even Higgs masses for $\tan\beta=10$, $M_0=1500$ GeV, $A_\kappa=-100$ GeV \label{fig:tanb10}}
\end{figure}

Here, we show some numerical results on the Higgs sector and fine-tuning.
(See in detail Refs.\cite{Kobayashi:2012ee,ours}.)
Figure \ref{fig:tanb10} shows  masses of the two light CP-even Higgs bosons, $h_1$ and $h_2$, as a function of $\lambda$ and $\kappa$ for $\tan\beta=10$ and $M_0 = 1500$ GeV. 
The red curve in the figure indicates the boundary where $\lambda$ and $\kappa$ blow up at the Planck scale. Inside this curve the model remains perturbative until the Planck scale. 
The yellow region is disfavored due to the false vacuum (see in detail \cite{Kanehata:2011ei}).
the lightest CP-even Higgs boson mass can not reach $125$ GeV. Instead, the second lightest CP-even Higgs boson mass can be $125$ GeV. 
In this region, the coupling squared to the gauge boson indicates that $h_1$ almost consists of the singlet and hence $h_2$ is almost doublet. 
For small $\tan \beta$, e.g. $\tan \beta =3$, there is a parameter region, where the lightest Higgs boson has $125$ GeV mass, too.

\begin{figure}[h]
\begin{center}
\begin{tabular}{l @{\hspace{10mm}} r}
\includegraphics[height=65mm]{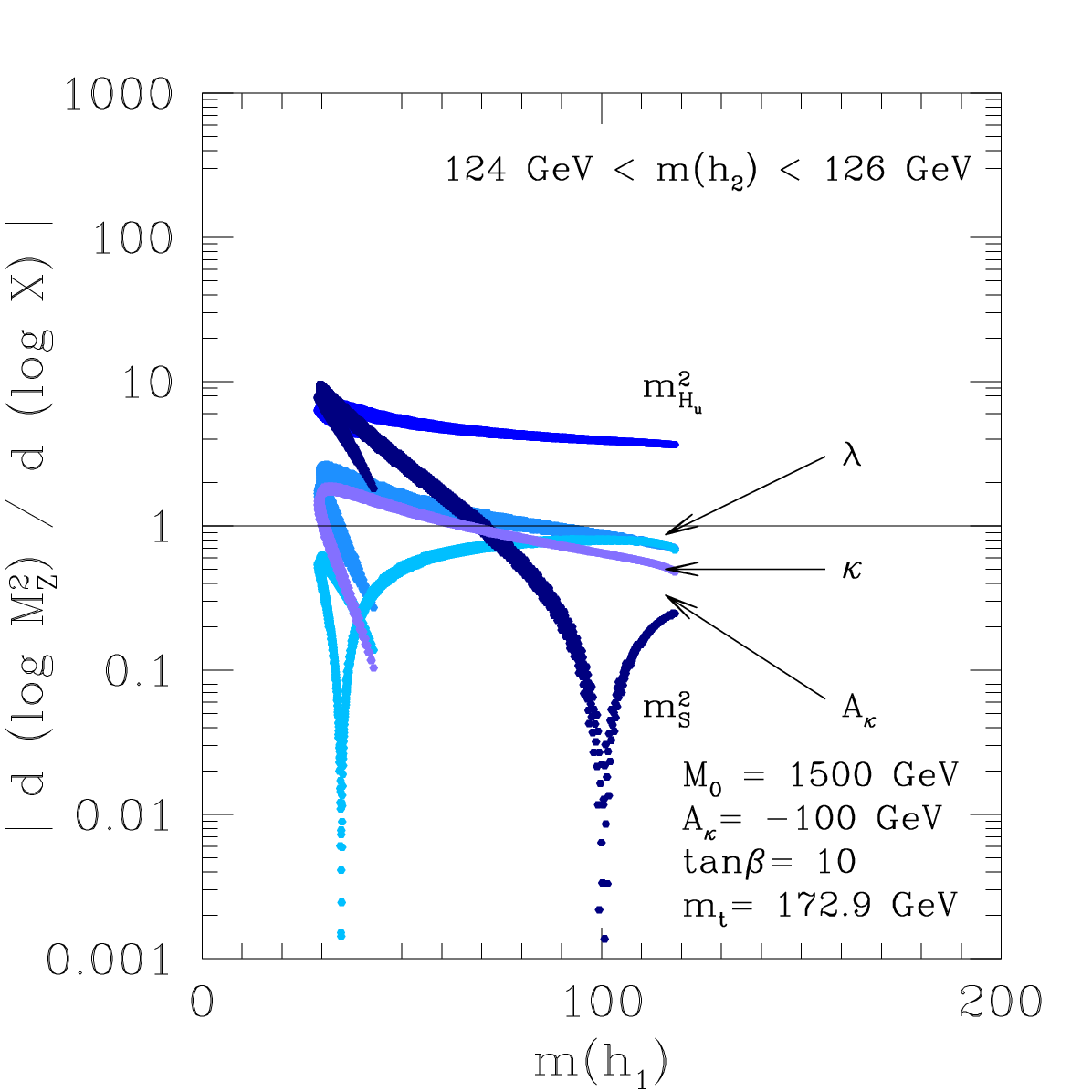} &
\includegraphics[height=65mm]{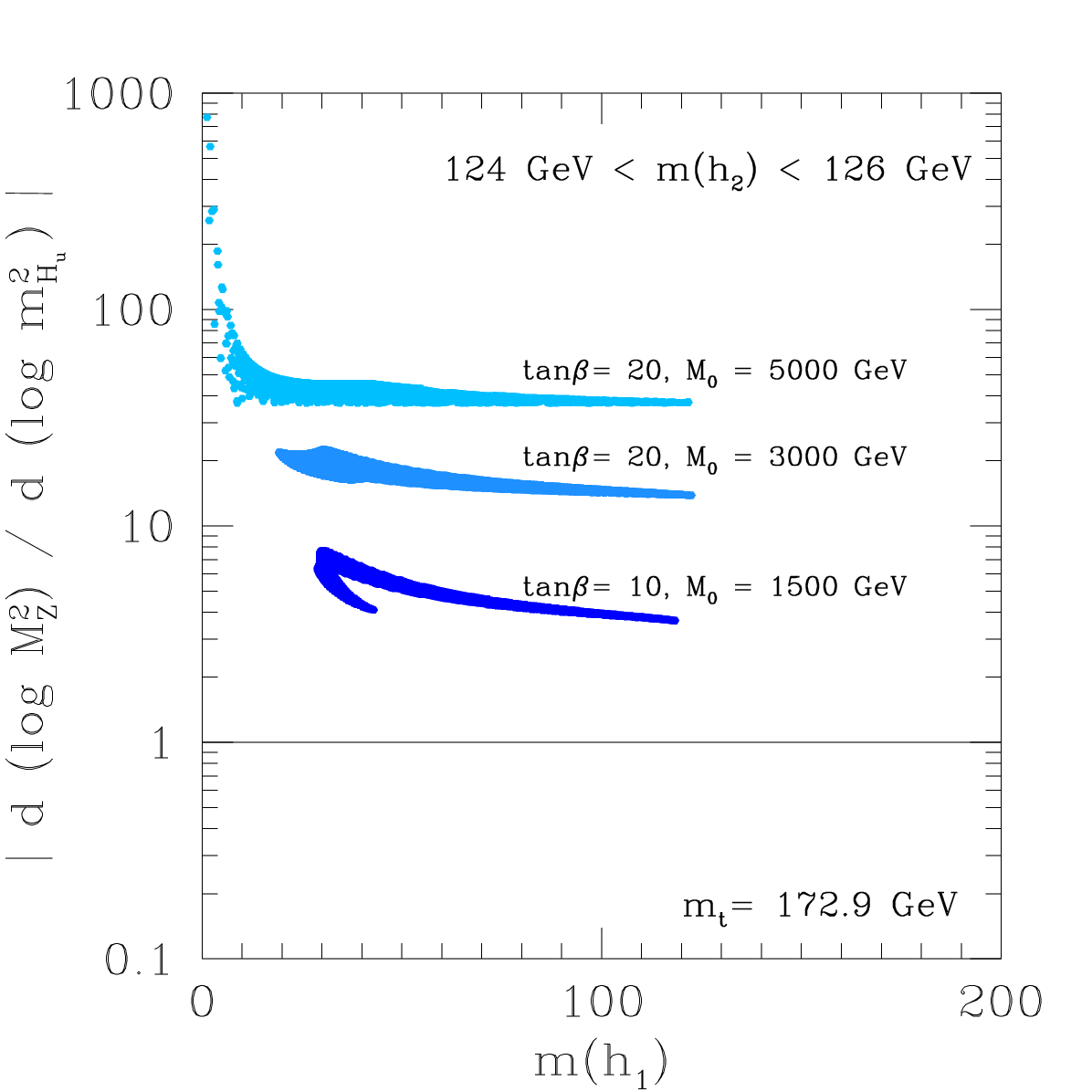}
\end{tabular}
\end{center}
\caption{The fine-tuning measures for $\tan\beta=10$, $A_\kappa=-100$ GeV.} \label{fig:FT-tanb10}
\end{figure}

We numerically estimate the degree of fine-tuning of the electroweak symmetry breaking in our model. Following the standard lore, we define the fine-tuning measure of a observable $y$ against an input parameter $x$ as,
\begin{equation}
\Delta_x^y = \frac{\partial \ln(y)}{\partial \ln(x)}.
\end{equation}
To evaluate the electroweak symmetry breaking, we set $y=M_Z^2$ and as $x$, we take $\lambda$, $\kappa$ and the small parameters, $m_{H_u}$, $m_S$ and $A_\kappa$ at the SUSY scale. 
Note that the large parameters such as $m_{H_d}$, $A_\lambda$ are not free parameters in our model and fixed by the ultraviolet physics. 
Figure \ref{fig:FT-tanb10} shows  the  fine-tuning measures $\Delta_{x}^{M_Z^2}$ for  the $\tan\beta=10$ case in the parameter region leading to
$124 {\rm GeV} < m_{h_2} < 126 {\rm GeV}$. 
In the left panel, we take $M_0=1500 $ GeV and $A_\kappa = -100$ GeV.
The fine-tuning measures $\Delta_{x}^{M_Z^2}$ is of ${\cal O}(10)$ or smaller.
The worst value is obtained for $\Delta_{x}^{M_Z^2}$ with $x=m_{H_u}^2$.
However, the severe fine-tining is not required, but ${\cal O}(10)\%$ tuning is enough for any parameter $x$.
The right panel shows  $\Delta_{m_{H_u}^2}^{M_Z^2}$ for  $M_0=3$ TeV and $M_0=5$ TeV ($\tan\beta=20$) compared with $M_0=1.5$ TeV ($\tan\beta=10$) case.
It is remarkable that only ${\cal O}(1)\%$ fine-tuning is required even for such a heavy spectrum.
That is, even the $5$ TeV case is acceptable in the standard of the conventional models build around $1$ TeV such as the constrained MSSM.

For small $\tan \beta$,  e.g. $\tan \beta =3$, we can realize a simliar value of fine-tuning measures,  $\Delta_{x}^{M_Z^2} \leq {\cal O}(10)$ 
in the parameter region leading to $124 {\rm GeV} < m_{h_1} < 126 {\rm GeV}$ for the lightest Higgs mass.
(See in detail Refs.\cite{Kobayashi:2012ee,ours}.)

All of three gaugino masses are equal to $M_0$ and heavy, and squark and slepton masses are similarly heavy.
The value of $\mu$ can vary from 100 GeV to about 500 GeV without introducing further fine-tuning.
The singlino mass is similar to@$\mu$.
Then, their mixing would be the lightest supersymmetric particle.

\section{Conclusion}

We have studied the NMSSM in the TeV scale mirage mediation.
We can realize the 125 GeV Higgs mass without severe fine-tuning.
${\cal O}(10)\%$ tuning is sufficient in our model, where 
cancellation between the first and the second terms in Eq.(\ref{eq:app-3})
is significant.
The size of $\mu$ is insensitive to $m_Z$, and it can be of ${\cal O}(400)$ GeV without introducing further fine-tuning.

\begin{acknowledgments}
The author would like to thank K.Hagimoto, H.Makino, K.Okumura, T.Shimomura and T.Takahashi.
This talk is based on the collaborations with them.
He is supported in part by a Grant-in-Aid for Scientific Research No.~25400252  from the Ministry of 
Education, Culture, Sports, Science and Technology of Japan. 
\end{acknowledgments}

\bigskip 

\end{document}